\documentclass[11pt,a4paper]{article} \usepackage{jheppub}

\newcommand{\lsim}
{\;\raisebox{-.3em}{\small $\stackrel{\displaystyle <}{\sim}$}\;}

\title{
Quantifying quark mass effects at the LHC: \\A study of $\bf{pp \to b \bar{b} 
b \bar{b} + X}$ at next-to-leading order }

\author[]{G. Bevilacqua, M. Czakon, M. Kr\"amer, M. Kubocz and M. Worek} 

\affiliation[]{Institute for Theoretical Particle Physics and Cosmology, 
RWTH Aachen University, D-52056 Aachen, Germany} 
\emailAdd{bevilacqua@physik.rwth-aachen.de}
\emailAdd{mczakon@physik.rwth-aachen.de}
\emailAdd{mkraemer@physik.rwth-aachen.de}
\emailAdd{kubocz@physik.rwth-aachen.de}  
\emailAdd{worek@physik.rwth-aachen.de} 

\abstract{ The production of four bottom quarks is an important
benchmark channel for Higgs analyses and searches for new physics at
the LHC.  We report on the calculation of the next-to-leading order
QCD corrections to the process  $pp \to b \bar{b} b \bar{b} + X$ with
the \textsc{Helac-NLO} automated framework, and present results for
inclusive cross sections and differential distributions.  We discuss
the impact of the higher-order corrections and, in particular, the
effect of the bottom quark mass. In addition, we provide an estimate of
the theoretical uncertainty from the variation of the renormalisation
and factorisation scales and the parton distribution functions. The
results are obtained with a new subtraction formalism for real
radiation at next-to-leading order,  implemented in the
\textsc{Helac-Dipoles} package. }

\dedicated{\rm TTK-13-11}

\begin{document} 
\maketitle
\flushbottom

%

\section{Introduction} 

The production of four bottom quarks, $pp \to b
\bar{b} b \bar{b} + X$, is an important background to various Higgs
analyses and new physics searches at the LHC, including for example
Higgs-boson pair production in two-Higgs doublet models at large
$\tan\beta$~\cite{Dai:1996rn}, or so-called hidden valley scenarios
where additional gauge bosons can decay into bottom
quarks~\cite{Strassler:2006im}. Accurate theoretical predictions for
the Standard Model production of multiple bottom quarks are thus
mandatory to exploit the potential of the LHC for new physics
searches. Furthermore, the calculation of the next-to-leading order (NLO)
QCD corrections to $pp \to b \bar{b} b \bar{b} + X$ provides a
substantial technical challenge and requires the development of
efficient techniques, with a high degree of automation.  We have
performed a NLO calculation of $ b \bar{b} b \bar{b} $ production at
the LHC with the \textsc{Helac-NLO} system~\cite{Bevilacqua:2011xh}. In
particular, we present results based on a new subtraction
formalism~\cite{Chung:2010fx, Chung:2012rq} for treating real
radiation corrections, as implemented in the \textsc{Helac-Dipoles}
package~\cite{Czakon:2009ss}. Two calculation  schemes have been
employed, the so-called four-flavour scheme (4FS) with only gluons and
light-flavour quarks in the proton, where massive bottom quarks are
produced from gluon splitting at short distances, and the
five-flavour-scheme (5FS)~\cite{Barnett:1987jw} with massless bottom
quarks as partons in the proton. At all orders in perturbation theory,
the four- and five-flavour schemes are identical, but the way of
ordering the perturbative expansion is different, and at any finite
order the results do not match. Comparing the predictions of the two
schemes at NLO thus provides a way to assess the theoretical uncertainty
from unknown higher-order corrections, and to study the effect of the
bottom mass on the inclusive cross section and on differential
distributions. First NLO results for $pp \to b \bar{b} b \bar{b} + X$
in the 5FS have been presented in Ref.\,\cite{Greiner:2011mp}. We not
only provide an independent calculation of this challenging process
with a different set of methods and tools, but also a systematic study
of the bottom quark mass effects by comparing the 5FS and 4FS
results. We note that NLO results for the production of four top quarks 
in hadron collisions have been discussed in Ref.\,\cite{Bevilacqua:2012em}.
In addition, NLO calculations for processes of similar complexity  
have recently been presented in the literature, including NLO QCD corrections to 
$t\bar{t}$ production in association with two jets \cite{ttjets}, 
the production of a single gauge boson plus jets \cite{boson},
double gauge boson production with two jets \cite{bosons}
and multi-jet production \cite{jets}.

The paper is organised as follows. In section~\ref{sec:setup} 
we briefly summarise
the set-up of the calculation. Numerical results for inclusive cross
sections and differential distributions for $ b \bar{b} b \bar{b} $
production at the LHC are presented in section~\ref{sec:num}. We
summarise in section~\ref{sec:conclusion}.

\section{Theoretical Framework}\label{sec:setup}
%
The calculation of the process $pp \to b \bar{b} b \bar{b} + X$ at NLO
QCD comprises the parton processes $gg \to b \bar{b} b \bar{b}$ and
$q\bar{q} \to b \bar{b} b \bar{b}$ at tree-level and including
one-loop corrections, as well as the tree-level parton processes $gg
\to b \bar{b} b \bar{b}+g$,  $q\bar{q} \to b \bar{b} b \bar{b} + g$,
$gq \to b \bar{b} b \bar{b} + q$ and $g\bar{q} \to b \bar{b} b \bar{b}
+ \bar{q}$. In the four-flavour scheme $q \in \{u,d,c,s\}$, and the
bottom quark is treated massive. The bottom mass effects are in
general suppressed by powers of $m_{b}/\mu$, where $\mu$ is the hard
scale of the process, e.g.\ the transverse momentum of a
bottom-jet. Potentially large logarithmic corrections $\propto
\ln(m_b/\mu)$ could arise from nearly collinear splitting of
initial-state gluons into bottom quarks, $g \to b\bar{b}$, where the
bottom mass acts as a regulator of the collinear singularity. This
class of $\ln(m_{b}/\mu)$-terms can be summed to all orders in
perturbation theory by introducing bottom parton densities in the
five-flavour scheme. The 5FS is based on the approximation that the
bottom quarks from the gluon splitting are produced at small
transverse momentum. However, in our calculation  we require that all
four bottom quarks can be experimentally detected, and we thus impose
a lower cut on the bottom transverse momentum, $p_{T,b} \ge
p_{T,b}^{\rm min}$. As a result, up to NLO accuracy the potentially
large logarithms in the process $pp \to b \bar{b} b \bar{b} + X$ are
replaced by $\ln(m_{b}/\mu) \to \ln(p_{T, b}^{\rm min}/\mu)$, with
$m_{b} \ll p_{T, b}^{\rm min} \lsim \mu$,  and are thus much less
significant numerically. Therefore, for the process at hand, the
differences between the 4FS and 5FS calculations with massive and
massless bottom quarks, respectively, should be moderate, but may not
be completely negligible. 

Our calculation is performed with the automated \textsc{Helac-NLO}
framework~\cite{Bevilacqua:2011xh}, which includes
\textsc{Helac-1loop}~\cite{vanHameren:2009dr} for the evaluation of
the numerators of the loop integrals and the rational terms,
\textsc{CutTools}~\cite{Ossola:2007ax}, which implements the OPP
reduction method~\cite{Ossola:2006us,Ossola:2008xq,Mastrolia:2008jb,
Draggiotis:2009yb}  
to compute one-loop amplitudes,
and \textsc{OneLoop}~\cite{vanHameren:2010cp} for the evaluation of
the scalar integrals. The singularities for soft and collinear parton
emission are treated using subtraction schemes as implemented in
\textsc{Helac-Dipoles}~\cite{Czakon:2009ss}, see the discussion below. The phase space
integration is performed with the help of the Monte Carlo generators
\textsc{Helac-Phegas}~\cite{Kanaki:2000ey,Papadopoulos:2000tt,Cafarella:2007pc}
and \textsc{Kaleu}~\cite{vanHameren:2010gg}, including
\textsc{Parni}~\cite{vanHameren:2007pt} for the importance sampling. 

The \textsc{Helac-Dipoles} package has been based on the standard
Catani-Seymour dipole subtraction formalism
\cite{Catani:1996vz,Catani:2002hc}. We have now extended
\textsc{Helac-Dipoles} by implementing a new subtraction
scheme~\cite{Chung:2010fx,Chung:2012rq} using the momentum mapping and
the splitting functions derived in the context of an improved parton
shower formulation by Nagy and Soper~\cite{Nagy:2007ty}. Compared to
standard dipole subtraction, the new scheme features a significantly
smaller number of subtraction terms and facilitates the matching of
NLO calculations with parton showers including quantum
interference. The results presented here constitute the first
application of the Nagy-Soper subtraction scheme for a $2 \to 4$
scattering process with massive and massless fermions. A detailed
description of the implementation of the new scheme, and a comparative
study of the numerical efficiency and the speed will be presented
elsewhere.  

%
\section{Numerical Results for the LHC}\label{sec:num}
%
In this section we present cross-section predictions for the process
$pp \to b \bar{b} b \bar{b} + X$ at the LHC at the centre-of-mass
energy of ${\sqrt{s} = 14}$\,TeV. We discuss the impact of the NLO-QCD
corrections, and study the dependence of the results on the bottom
quark mass. We show results obtained with the standard dipole
subtraction scheme for real radiation, and with the new subtraction
formalism~\cite{Chung:2010fx, Chung:2012rq} as implemented in
\textsc{Helac-Dipoles}. 

Let us first specify the input parameters and the acceptance cuts we
impose. The top quark mass, which appears in the loop corrections, is
set to $m_{t} = 173.5$\,GeV~\cite{Beringer:1900zz}.  We combine
collinear final-state partons with pseudo-rapidity $|\eta| <5$ into
jets according to the anti-$k_T$ algorithm~\cite{Cacciari:2008gp} with
separation $R = 0.4$. The bottom-jets have to pass the transverse
momentum and rapidity cuts $p_{T,b} > 30$\,GeV and $|y_{b}|< 2.5$,
respectively.  The renormalisation and factorisation scales are set to
the scalar sum of the bottom-jet transverse masses, $\mu_{R} = \mu_{F} =
\mu_0 = H_{T}$, with $H_{T} =
m_{T,b}+m_{T,\bar{b}}+m_{T,b}+m_{T,\bar{b}}$ and the transverse mass
$m_{T,b}=\sqrt{m^2_{b}+p^2_{T,b}}$. For the five-flavour scheme
calculation with massless bottom quarks the transverse mass equals the
transverse momentum, $m_{T,b} = p_{T,b}$.  Note that the implementation 
of a dynamical scale requires a certain amount of care, as the subtraction 
terms for real radiation have to be evaluated with a different kinematical 
configuration specified by the momentum mapping of the subtraction scheme. 
Comparing our results as obtained with the 
Catani-Seymour subtraction and the Nagy-Soper scheme, which is based on 
a different momentum mapping, provides an important and highly non-trivial 
internal check of our calculation. 

\subsection{Massless bottom quarks within the five-flavour scheme}

Results are presented for the NLO CT10~\cite{Lai:2010vv} and
MSTW2008~\cite{Martin:2009iq} parton distribution functions (pdfs)
with five active flavours and the corresponding two-loop $\alpha_{\rm
s}$. To study the impact of the higher-order corrections, we also show
leading-order results obtained using the CT09MC1~\cite{Lai:2009ne} and
MSTW2008  LO pdf sets and one-loop running for $\alpha_{\rm s}$. 

We first discuss the impact of the bottom-quark induced processes,
$b\bar{b} \to b \bar{b} b \bar{b}$, on the hadronic cross section at
leading-order. The difference between the $q\bar{q}$ initiated
processes,  $q\bar{q} \to b \bar{b} b \bar{b}$, with and without
bottom-quarks is at the level of $2.5\%$. Moreover, at the central
scale, $\mu = H_{T}$,  the hadronic cross section is completely
dominated by gluon-fusion, with only about 1\% contribution of all
quark-antiquark annihilation processes. The bottom-induced
contributions are thus negligible, and we decided to neglect bottom
initial states in the computation of the cross section both at LO and
NLO. Note that the suppression of the bottom-induced processes, which
include for example potentially large forward scattering of
bottom-quarks through $t$-channel gluon exchange, depends crucially on
the transverse momentum cuts we impose on the bottom-jets. 

\begin{figure}
\begin{center}
\includegraphics[width=0.8\textwidth]{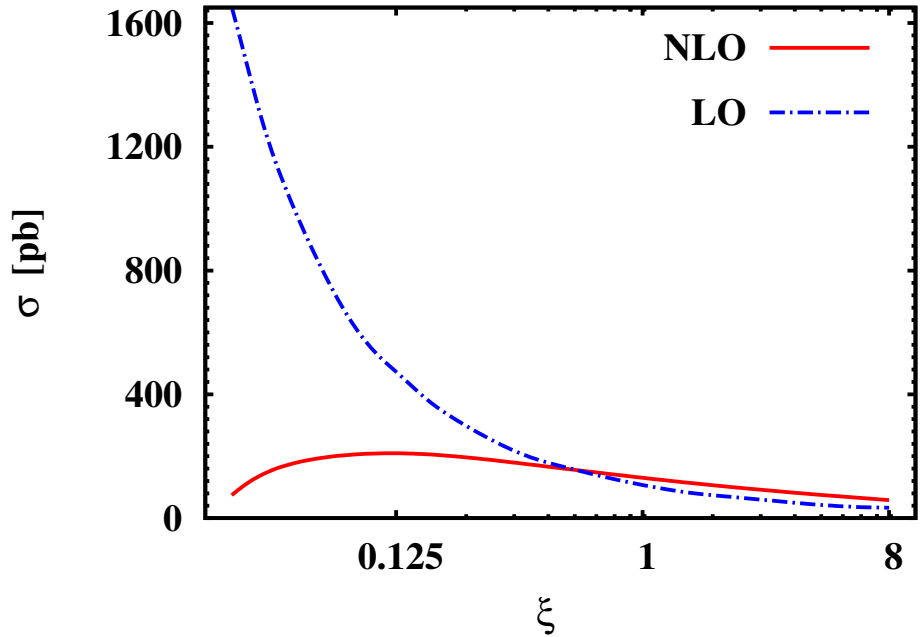}
\end{center}
\caption{\it \label{fig:scale-dependence} Scale dependence  of the 5FS LO
  and NLO cross sections for  $pp\rightarrow b\bar{b} b\bar{b} ~+
  X$ at the LHC ($\sqrt{s}$ = 14 TeV). The renormalisation and factorisation 
scales are set to a common value $\mu_R=\mu_F= \xi
  \, \mu_0$ where  $\mu_0= H_{T}$. The CT09MC1 and CT10 pdf sets have been used
 for the LO and NLO cross sections, respectively.}
  \end{figure}

Let us first present our results for the inclusive cross section
$pp\to b\bar{b}b\bar{b} +X$ at the LHC ($\sqrt{s}$ = 14 TeV),
including the transverse momentum and rapidity cuts specified at the
beginning of the section. The NLO QCD corrections strongly reduce the
scale dependence, as demonstrated in
Figure\,\ref{fig:scale-dependence}. The central cross section
predictions are collected in Table\,\ref{tab:5fs}. Varying the
renormalisation and factorisation scales simultaneously about the
central scale by a factor of two, we find a residual scale uncertainty
of approximately 30\% at NLO, a reduction by about a factor of two
compared to LO. The size of the $K$-factor, $K = \sigma_{\rm
NLO}/\sigma_{\rm LO}$, strongly depends on the pdf set, with $K =
1.15$ for CT10 and $K = 1.37$ for MSTW2008. We emphasise, however, that the
$K$-factor is an unphysical quantity and strongly sensitive to the
choice of scale through the large LO scale dependence.
\begin{table}[h!]
\renewcommand{\arraystretch}{1.5}
\begin{center}
  \begin{tabular}{c|c|c|c}
\hline
$pp\to b\bar{b}b\bar{b}+X$& $\sigma_{\rm LO}$\,[pb] 
& $\sigma_{\rm NLO}$\,[pb] 
& $K = \sigma_{\rm NLO}/\sigma_{\rm LO}$ \\ \hline \hline
CT09MC1/CT10 &  $106.9^{+61.5\,(57\%)}_{-36.4\,(34\%)}$ 
& $123.6^{+35.6\,(29\%)}_{-26.6\,(22\%)}$ 
& 1.15 \\
MSTW2008LO/NLO & $99.9^{+58.7\,(59\%)}_{-34.9\,(35\%)}$ 
& $136.7^{+38.8(28\%)}_{-30.9\,(23\%)}$ 
& 1.37\\
\hline
  \end{tabular}
\end{center}
  \caption{\it \label{tab:5fs} 5FS LO and NLO cross sections for
$pp\rightarrow b\bar{b} b\bar{b} ~+ X$ at the LHC ($\sqrt{s}$ = 14
TeV). The renormalisation and factorisation scales have been set to
the central value $\mu_0 = H_T$,  and the uncertainty is estimated by
varying both scales simultaneously by a factor two about the central
scale.  Results are shown for the CT09MC1/CT10 and MSTW2008LO/NLO pdf
sets.} 
 \end{table}

We observe a difference of about $-7\%$ and $+11\%$ between the CT10
and MSTW2008 pdf parametrisations at LO and NLO, respectively.  The pdf
uncertainty as estimated from the MSTW2008 error pdf
sets~\cite{Martin:2009iq} amounts to $+7.3\%$ and  $-1.5\%$ at 68\%
C.L., and is significantly  smaller than the scale uncertainty. A more
systematic discussion of pdf and $\alpha_{\rm s}$ uncertainties will
thus be referred to a forthcoming publication. 

\begin{figure}
\begin{center}
\includegraphics[width=0.49\textwidth]{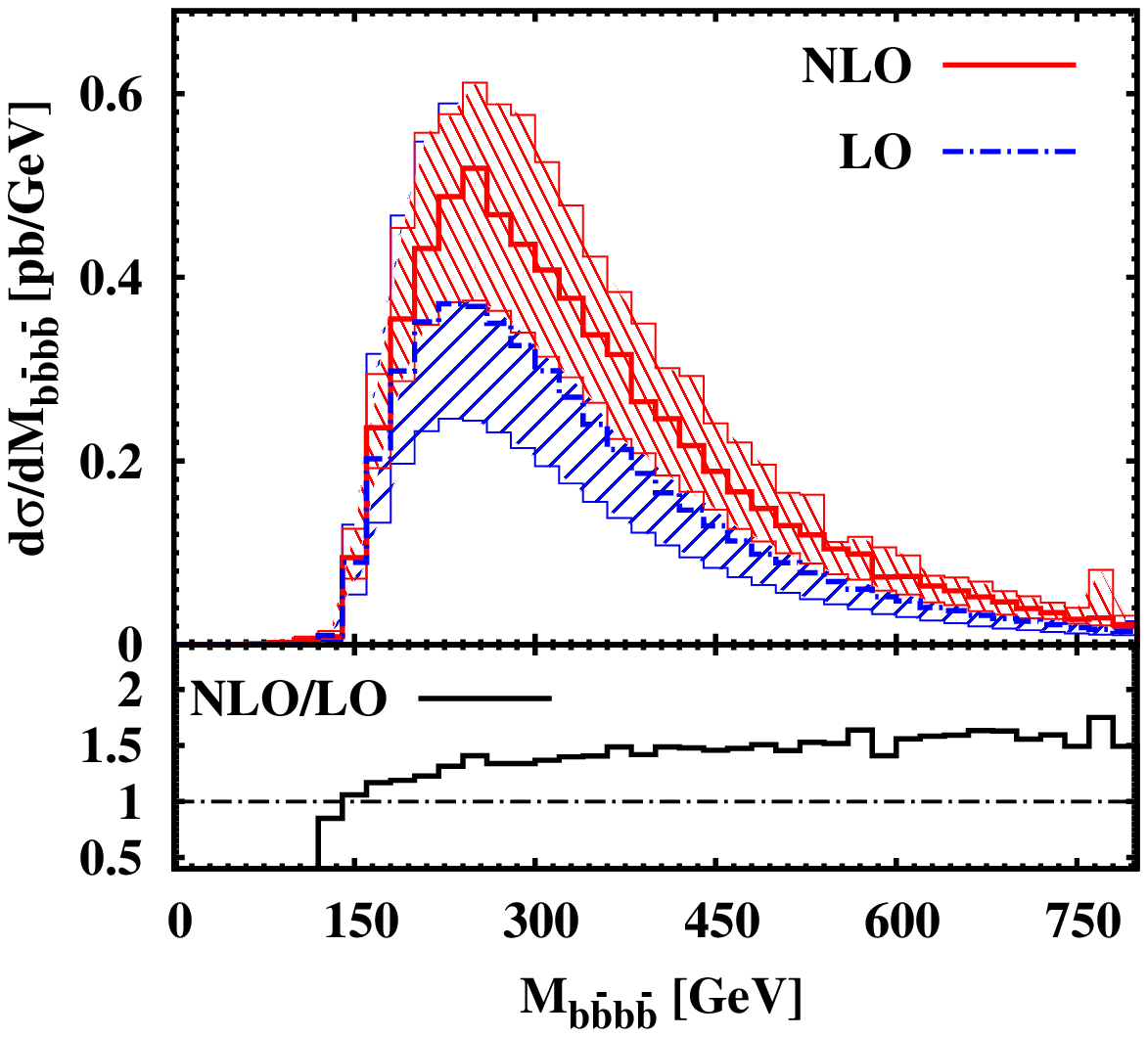}
\includegraphics[width=0.49\textwidth]{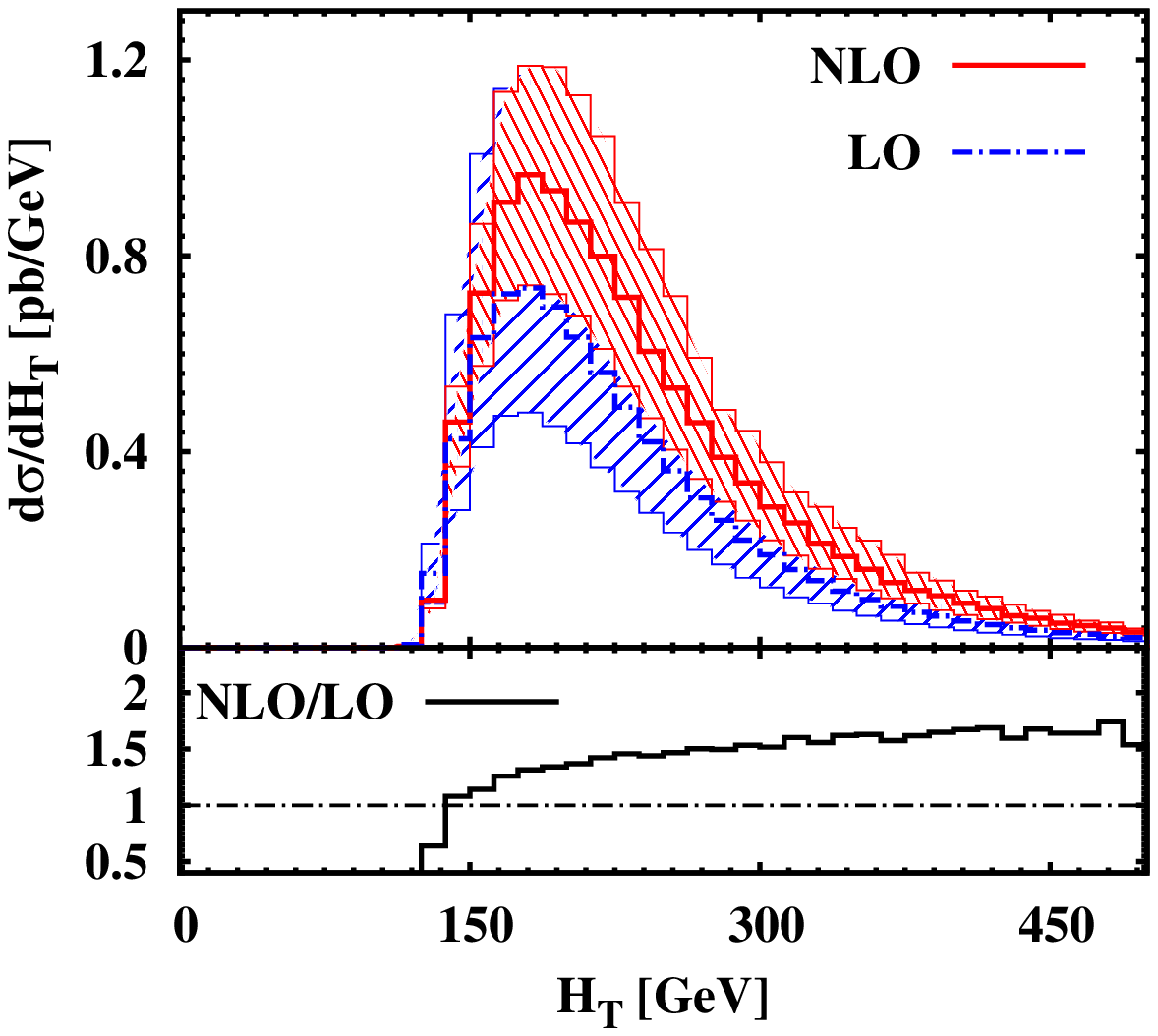}\\[5mm]
%
\includegraphics[width=0.49\textwidth]{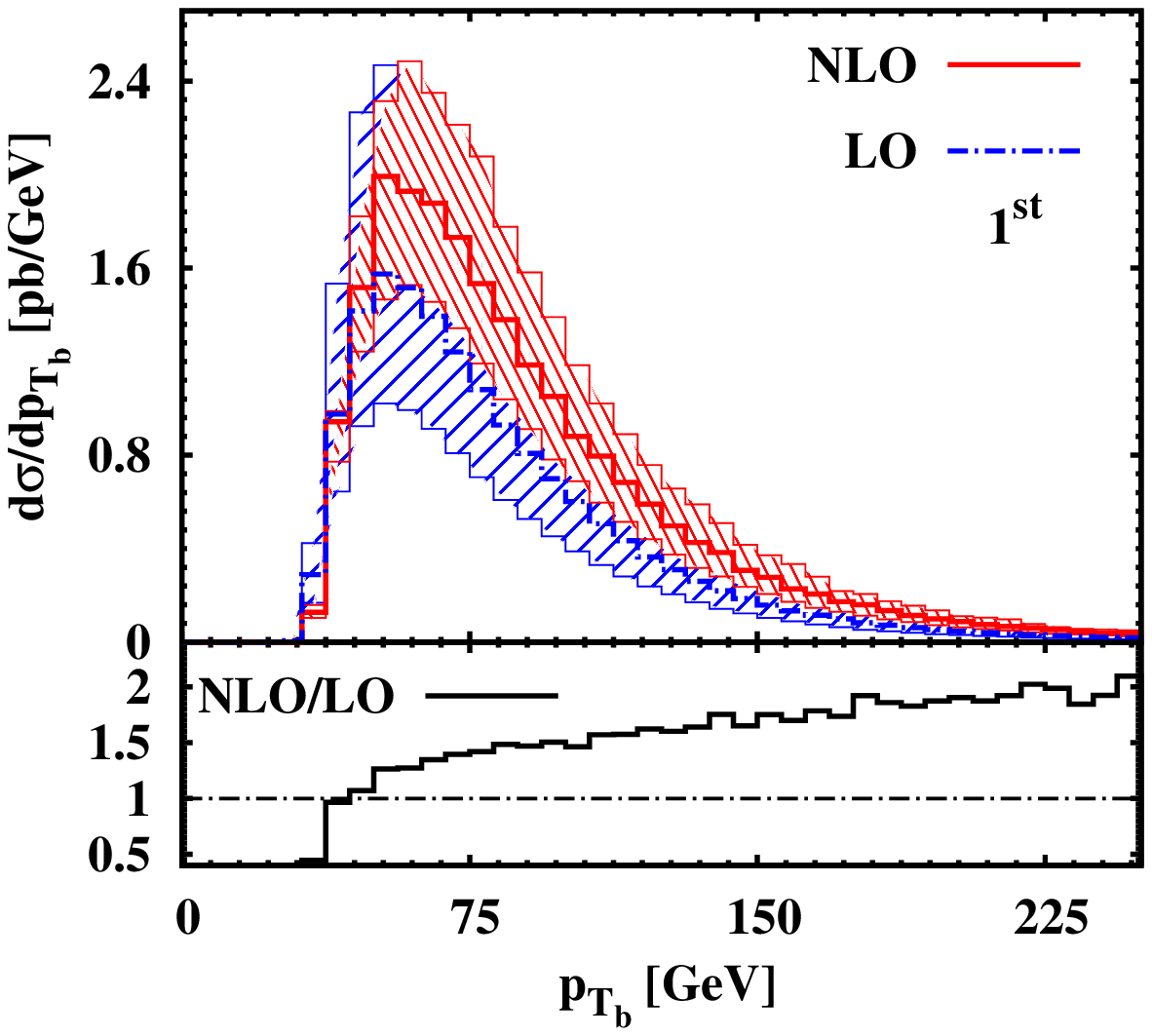}
\includegraphics[width=0.49\textwidth]{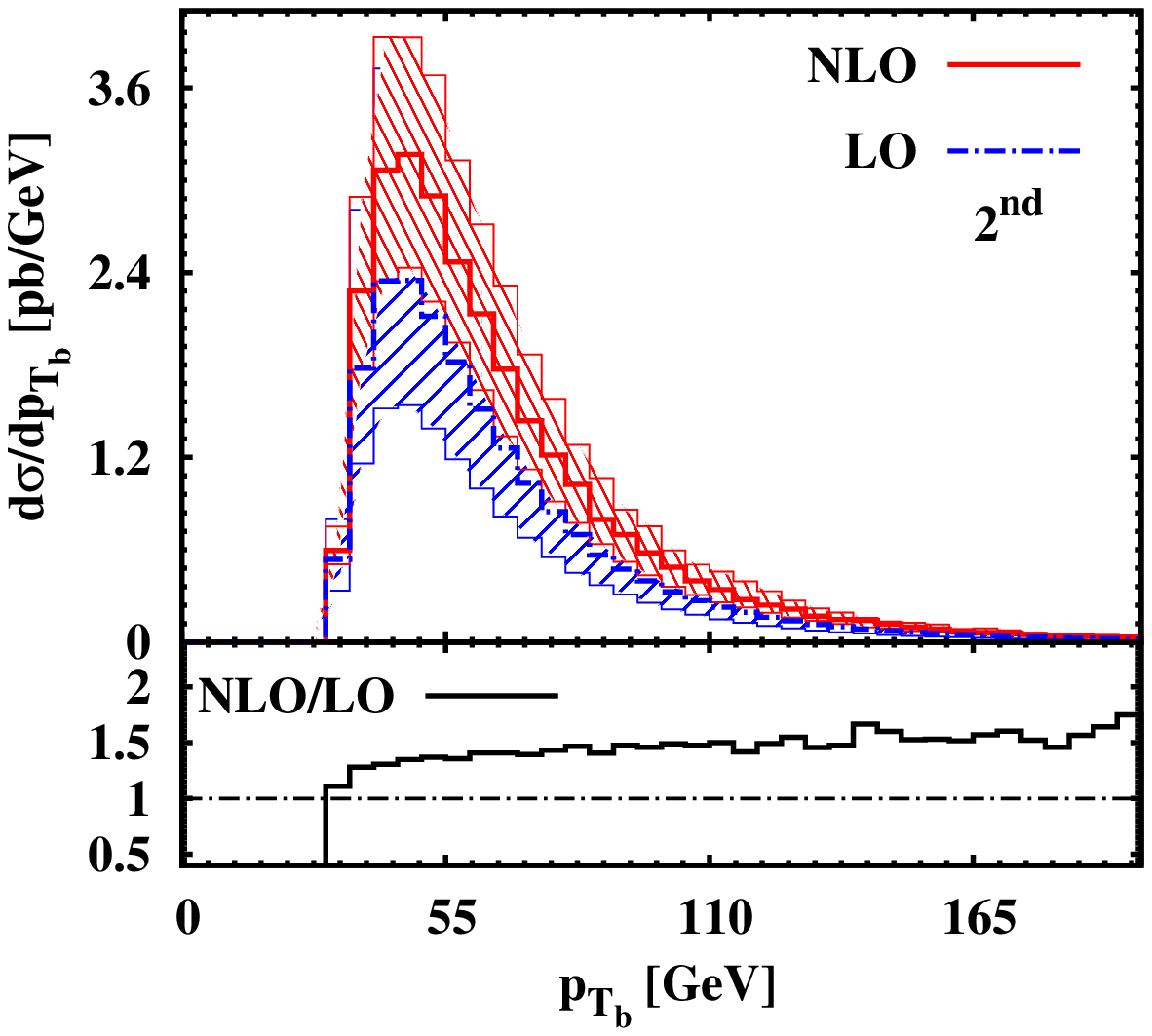}
\end{center}
\caption{\it \label{fig:nlo2:5f}    Differential cross section for
$pp\rightarrow b\bar{b} b\bar{b} ~+ X$ at the LHC ($\sqrt{s}$ = 14
TeV) in the 5FS as a function of the invariant mass of the
$b\bar{b}b\bar{b}$ system (upper left panel), the total transverse energy
of the system (upper right panel), the transverse momentum of the
hardest bottom jet (lower left panel) and the transverse momentum of the 
second hardest bottom jet (lower right panel). The
dash-dotted (blue) curve corresponds to the LO and  the solid (red)
curve to the NLO result. The scale choice is $\mu_R = \mu_F = \mu_0 =
H_T$. The hashed area represents the  scale uncertainty, and the lower
panels display the differential K factor. The cross sections  are
evaluated with the MSTW2008 pdf sets.}
\end{figure}

An important input for the experimental analyses and the
interpretation of the experimental data are accurate predictions of
differential distributions. Our calculation is set up as a
parton-level Monte Carlo program and thus allows us to predict any
infrared-safe observable at NLO. Figure~\ref{fig:nlo2:5f} shows LO and 
NLO predictions for the invariant mass of the $b\bar{b}b\bar{b}$ system 
(upper left panel), the total transverse energy $H_{T}$
(upper right panel),  the transverse momentum of
the hardest bottom jet (lower left panel) and the
transverse momentum of the second hardest bottom jet
(lower right panel). We also show the theoretical
uncertainty through scale variation and the $K$-factor as a function
of the kinematic variable. It is evident from
Figure~\ref{fig:nlo2:5f} that the NLO
corrections significantly reduce the theoretical uncertainty of the
differential distributions, and that the size of the higher-order
effects depends on the kinematics. For an accurate description of
exclusive observables and differential distributions it is thus not
sufficient to rescale a LO prediction with an inclusive $K$-factor.

As discussed in section\,\ref{sec:setup} we have performed the
calculation with two different subtraction schemes, the standard
Catani-Seymour (CS) dipole subtraction, and a new scheme based on the
splitting functions and momentum mapping of an improved parton shower
by Nagy and Soper (NS).  The comparison between the two schemes for
the inclusive 5FS cross section is presented in
Table\,\ref{tab:5fsns}. For the Catani-Seymour scheme we show results
without ($\alpha_{\rm max}=1$) and with ($\alpha_{\rm max}=0.01$) a
restriction on the phase space of the subtraction as proposed in
Ref.\,\cite{Nagy:1998bb,Nagy:2003tz}. As evident from Table\,\ref{tab:5fsns}, the
cross sections obtained using the Catani-Seymour (CS) and Nagy-Soper
(NS) subtraction schemes agree within the numerical uncertainty of the
Monte Carlo integration. This result not only provides a validation of
our implementation of the new subtraction scheme into
\textsc{Helac-Dipoles}, but also provides a non-trivial internal cross
check of the calculation. 
\begin{table}[h!]
\renewcommand{\arraystretch}{1.5}
\begin{center}
  \begin{tabular}{c|c|c|c}
\hline
$pp\to b\bar{b}b\bar{b}+X$  
& $\sigma_{\rm NLO}^{\rm CS \, (\alpha_{max}=1)}$ [pb] & 
$\sigma_{\rm NLO}^{\rm CS \, (\alpha_{max}=0.01)}$ [pb] & 
$\sigma_{\rm NLO}^{\rm NS}$ [pb] \\ \hline \hline
CT10 & $123.6 \pm 0.4$ & 
$124.9 \pm 0.9$ & $124.8 \pm 0.3$\\
MSTW2008NLO  & $136.7 \pm 0.3$ & 
$136.1 \pm 0.5$ & $137.6 \pm 0.5$\\
\hline
  \end{tabular}
\end{center}
  \caption{\it \label{tab:5fsns} 5FS NLO cross sections for
$pp\rightarrow b\bar{b} b\bar{b} ~+ X$ at the LHC ($\sqrt{s}$ = 14
TeV). Results are shown for two different subtraction schemes, the
Catani-Seymour (CS) dipole subtraction, without ($\alpha_{\rm max}=1$)
and with ($\alpha_{\rm max}=0.01$) a restriction on the phase space of
the subtraction, and the new Nagy-Soper (NS) scheme, including the
numerical error from the Monte Carlo integration. The renormalisation
and factorisation scales have been set to the central value $\mu_0 =
H_T$,  and the CT10 and MSTW2008NLO pdf sets have been
employed.} 
 \end{table}

\begin{figure}
\begin{center}
\includegraphics[width=0.49\textwidth]{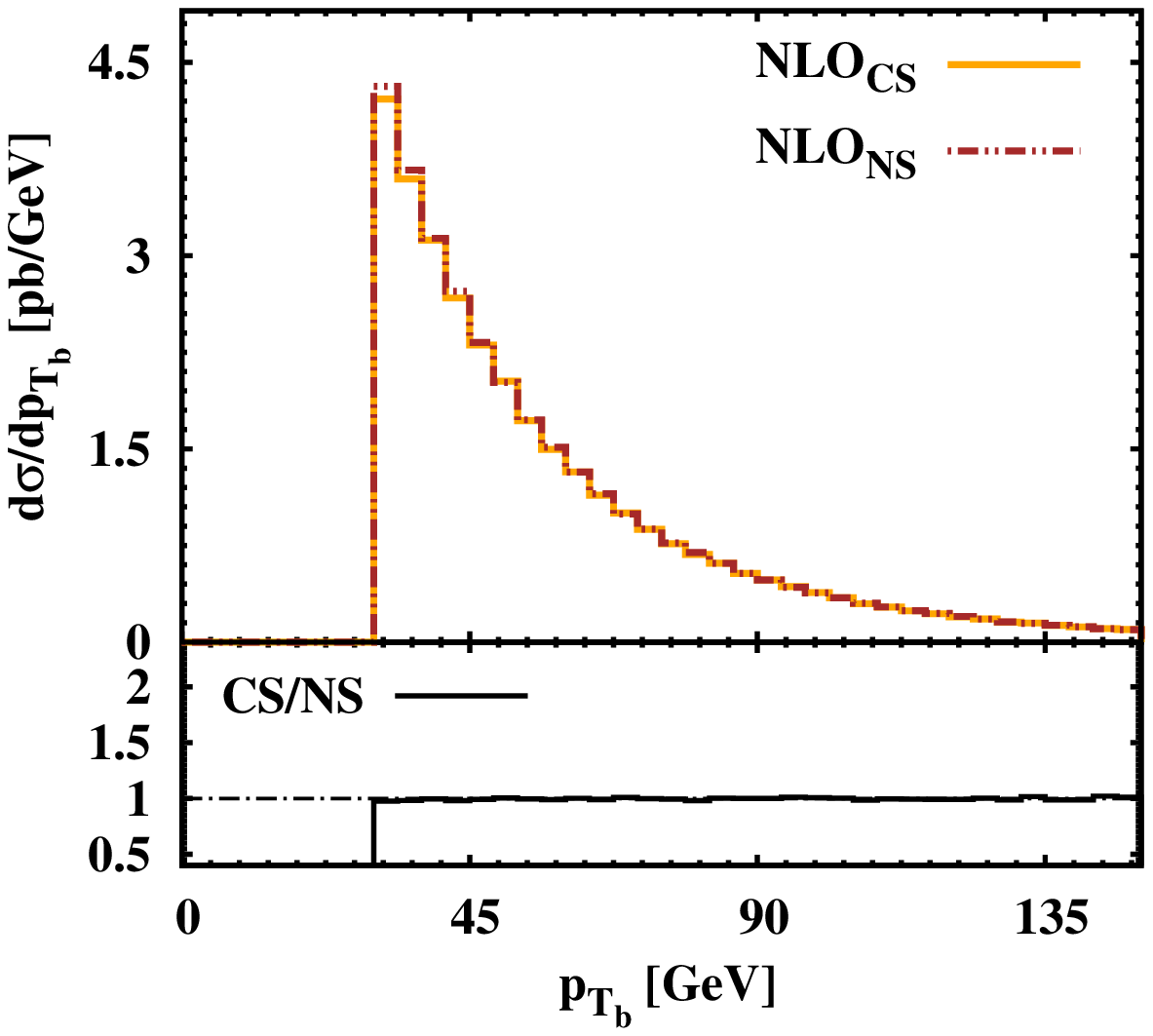}
\includegraphics[width=0.49\textwidth]{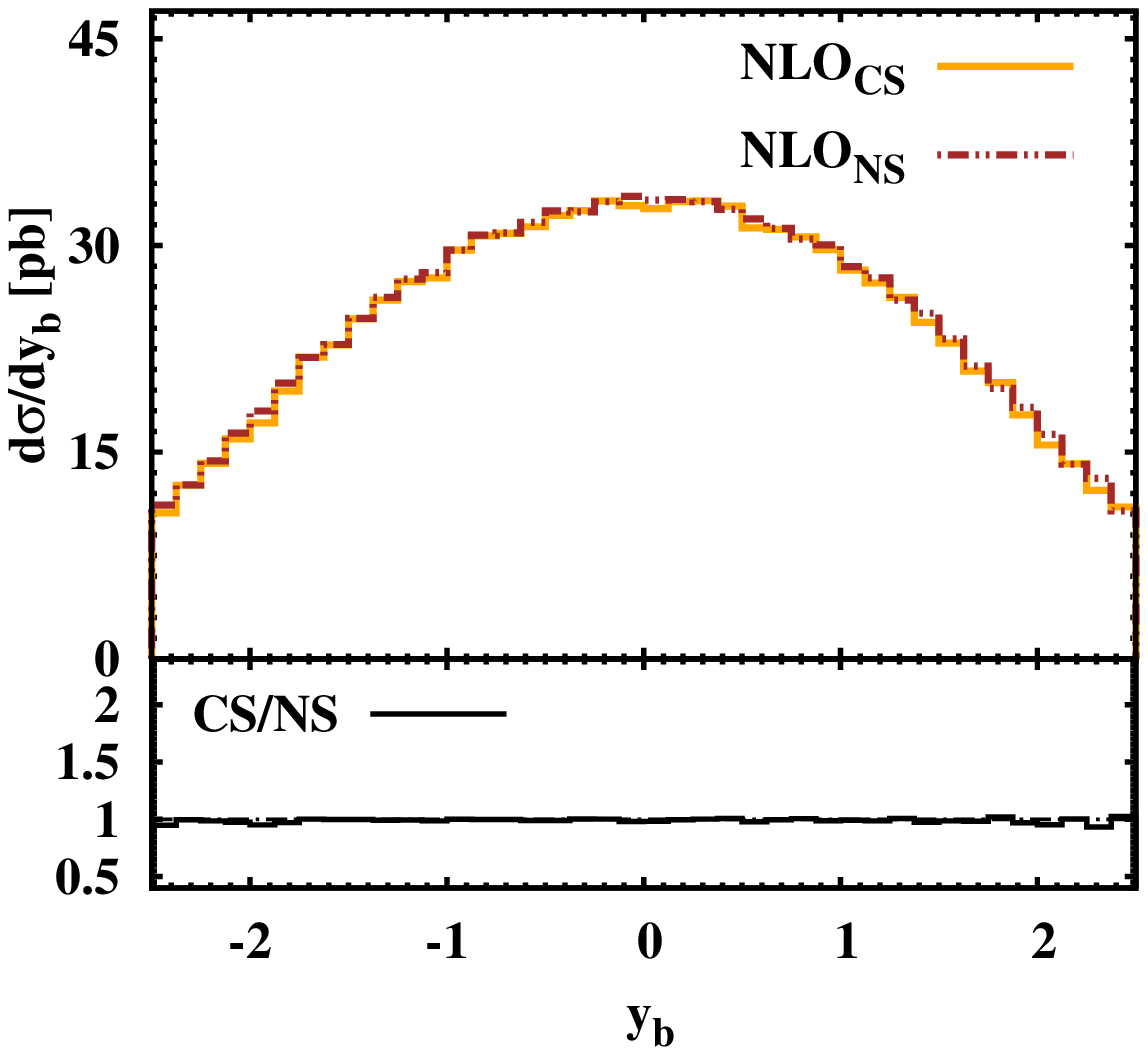}\\[5mm]
\includegraphics[width=0.49\textwidth]{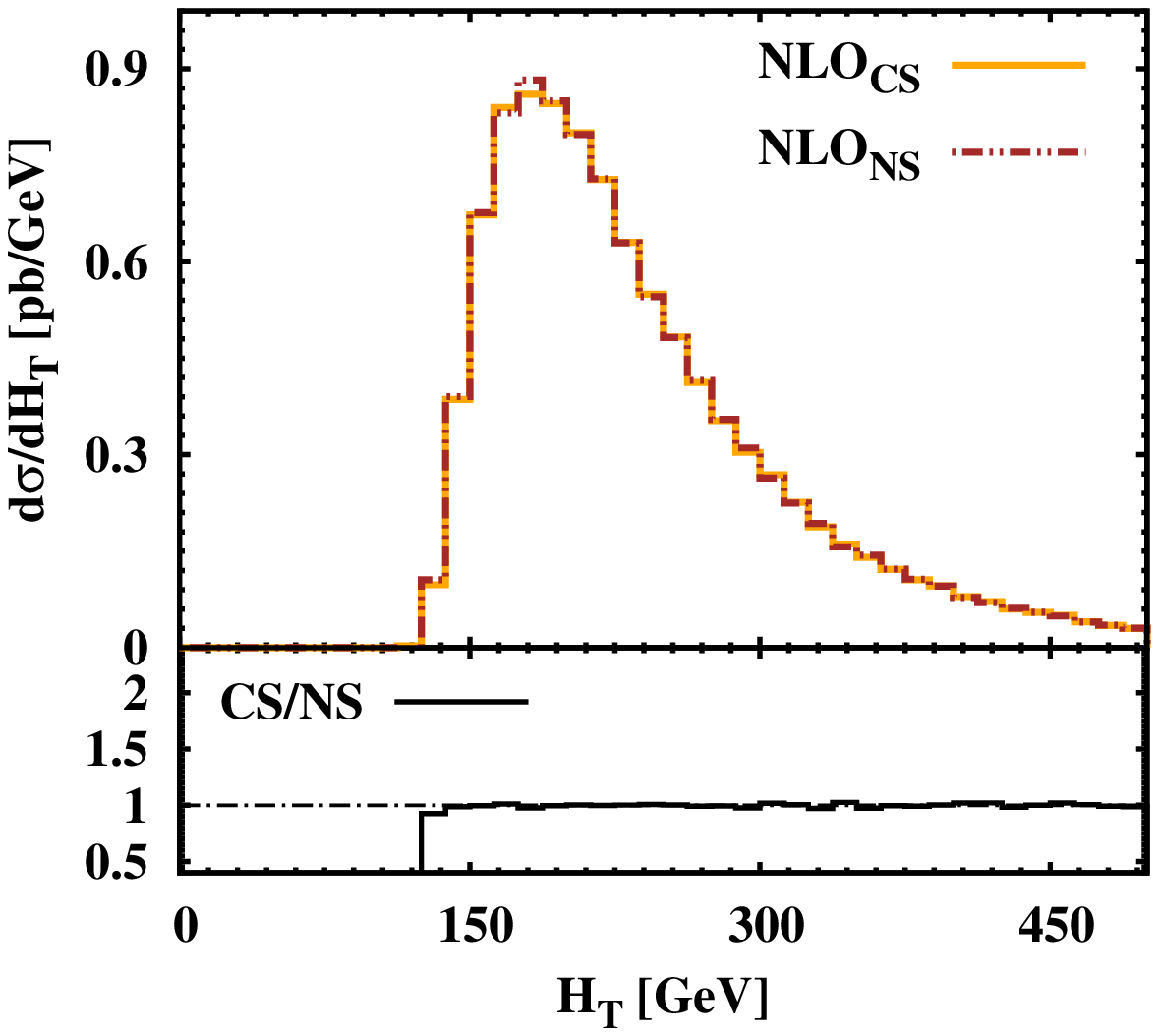}
\includegraphics[width=0.49\textwidth]{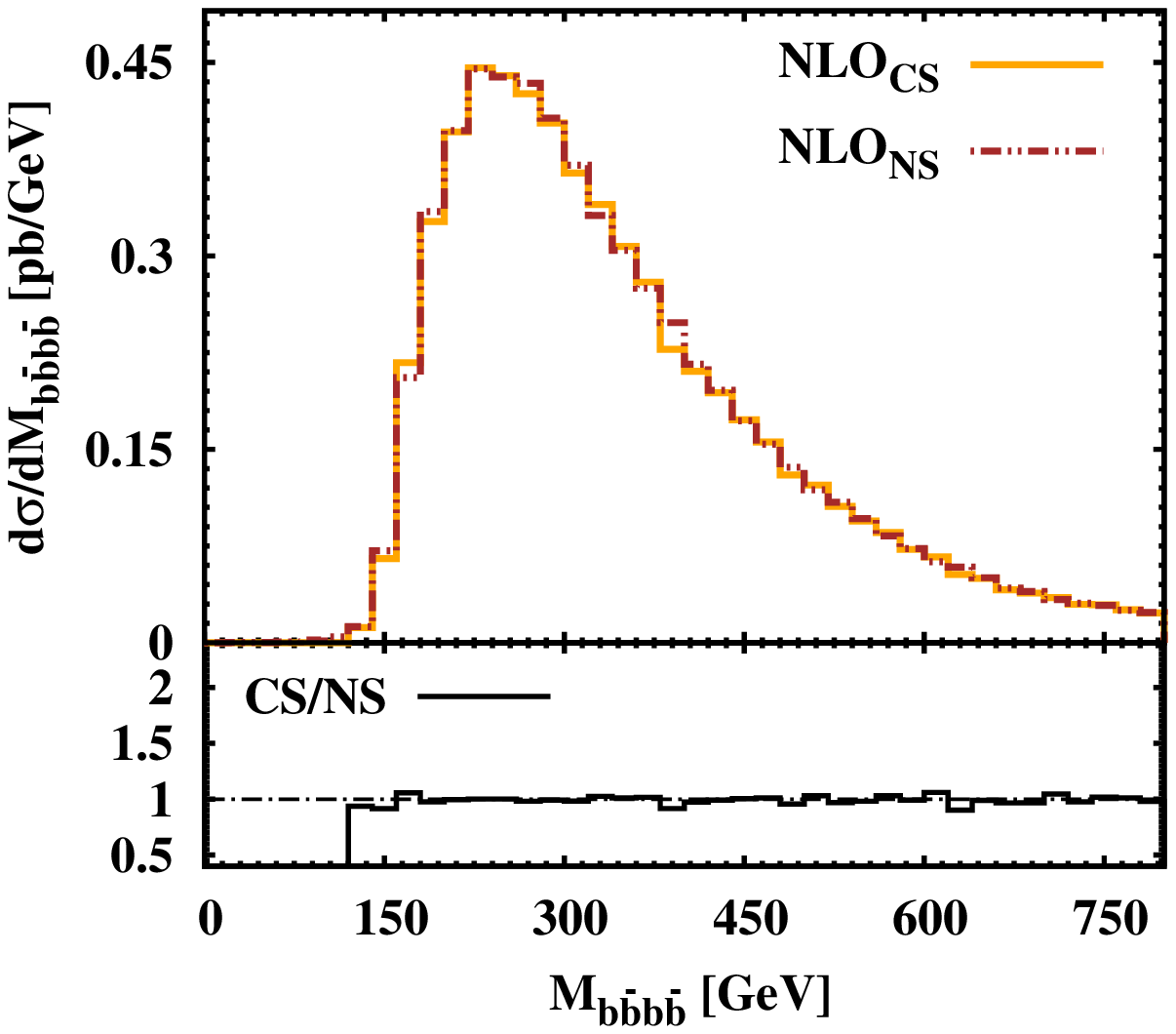}
\end{center}
\caption{\it \label{fig:CS-NS}   Differential cross section for
$pp\rightarrow b\bar{b} b\bar{b} ~+ X$ at the LHC ($\sqrt{s}$ = 14
TeV) in the 5FS as a function of the average transverse momentum 
of the bottom jets (upper left panel), the average rapidity of the
bottom jets (upper right panel), the total transverse energy (lower left panel) 
and the $b\bar{b}b\bar{b}$ invariant mass (lower right panel).  The
dash-dotted (brown) curve corresponds to the Catani-Seymour (CS) and the
solid (orange) curve to the Nagy-Soper (NS) subtraction schemes, respectively. The
lower panels show the ratio of the results within the two schemes.
The scale choice is $\mu_R = \mu_F = \mu_0 = H_T$, the cross sections
are evaluated with the CT10 pdf set.}
\end{figure}

We have also compared the results obtained in the CS and NS
subtraction schemes for various differential distributions.  Some
examples are collected in Figure\,\ref{fig:CS-NS}. We observe full
agreement between the predictions calculated with the two schemes.

\subsection{Massive bottom quarks within the four-flavour scheme}
%
Within the four-flavour scheme bottom quarks are treated massive and
are not included in the parton distribution functions of the proton.
We define the bottom quark mass in the on-shell scheme and use $m_b = 4.75$\,GeV, consistent with the choice made in the  MSTW2008
four-flavour pdf. The central cross section prediction in LO and NLO
for $\mu = H_T$ using the 4FS MSTW2008~\cite{Martin:2010db} pdf are
shown in Table\,\ref{tab:4fs}. Comparing with the 5FS results
presented in Table\,\ref{tab:5fs}, we observe that the bottom mass
effects decrease the cross section prediction by $18\%$ at LO and
$16\%$ at NLO. The residual scale dependence at NLO is approximately
30\%, similar to the 5FS calculation.
\begin{table}[h!]
\renewcommand{\arraystretch}{1.5}
\begin{center}
  \begin{tabular}{c|c|c|c}
\hline
$pp\to b\bar{b}b\bar{b}+X$& $\sigma_{\rm LO}$\,[pb] & 
$\sigma_{\rm NLO}$\,[pb] & $K = \sigma_{\rm NLO}/\sigma_{\rm LO}$ \\ \hline \hline 
MSTW2008LO/NLO (4FS) & $84.5^{+49.7(59\%)}_{-29.6(35\%)} $ & 
$118.3^{+33.3(28\%)}_{-29.0(24\%)}$ & 1.40\\
\hline
  \end{tabular}
\end{center}
  \caption{\it \label{tab:4fs} 4FS LO and NLO cross sections for
$pp\rightarrow b\bar{b} b\bar{b} ~+ X$ at the LHC ($\sqrt{s}$ = 14
TeV). The renormalisation and factorisation scales have been set to
the central value $\mu_0 = H_T$,  and the uncertainty is estimated by
varying both  scales
simultaneously by a factor two about the central scale.  Results are
shown for the 4FS MSTW2008LO/NLO pdf sets.} 
 \end{table}

The difference between the massless 5FS and the massive 4FS
calculations has two origins.  First, there are genuine bottom mass
effects, the size of which depends  sensitively on the transverse
momentum  cut. For $p_{T,b}^{\rm min} = 30$\,GeV we find a 10\%
difference between the 5FS and 4FS  from non-singular bottom-mass
dependent terms. This difference decreases to about $1\%$ for
$p_{T,b}^{\rm min} = 100$\,GeV. Second, the two calculations involve
different pdf sets and  different corresponding $\alpha_{\rm
  s}$. While a 4FS pdf has, in general, a larger gluon flux than a 5FS pdf, as
there is no $g\to b \bar{b}$ splitting, the corresponding four-flavour
$\alpha_{\rm s}$  is smaller than for five active flavours. For
$pp\rightarrow b\bar{b} b\bar{b} ~+ X$ the difference in $\alpha_{\rm
  s}$ is  prevailing and results in a further reduction of the 4FS cross
section prediction by about $5\%$.  This latter difference
should be viewed as a scheme dependence rather than a bottom mass
effect. 

\begin{figure}
\begin{center}
\includegraphics[width=0.49\textwidth]{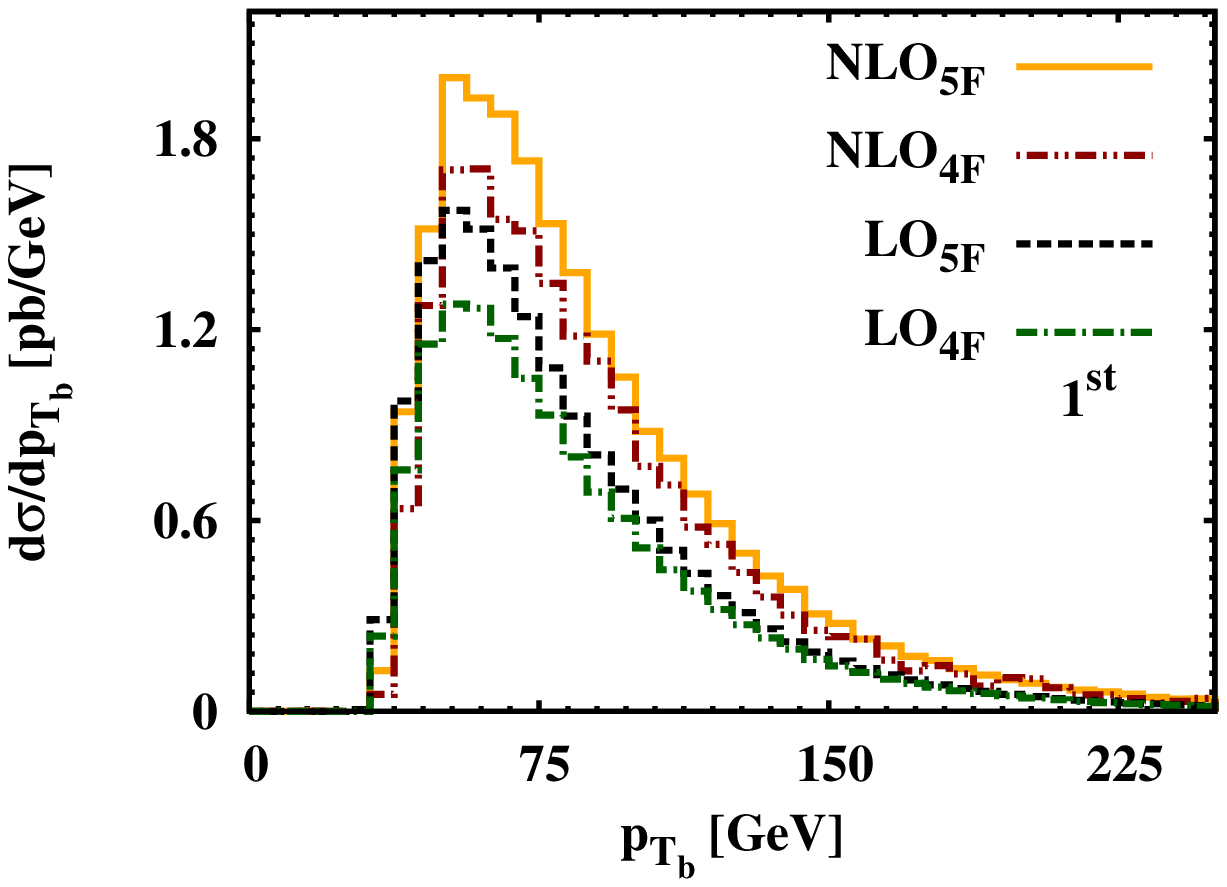}\\[5mm]
\includegraphics[width=0.49\textwidth]{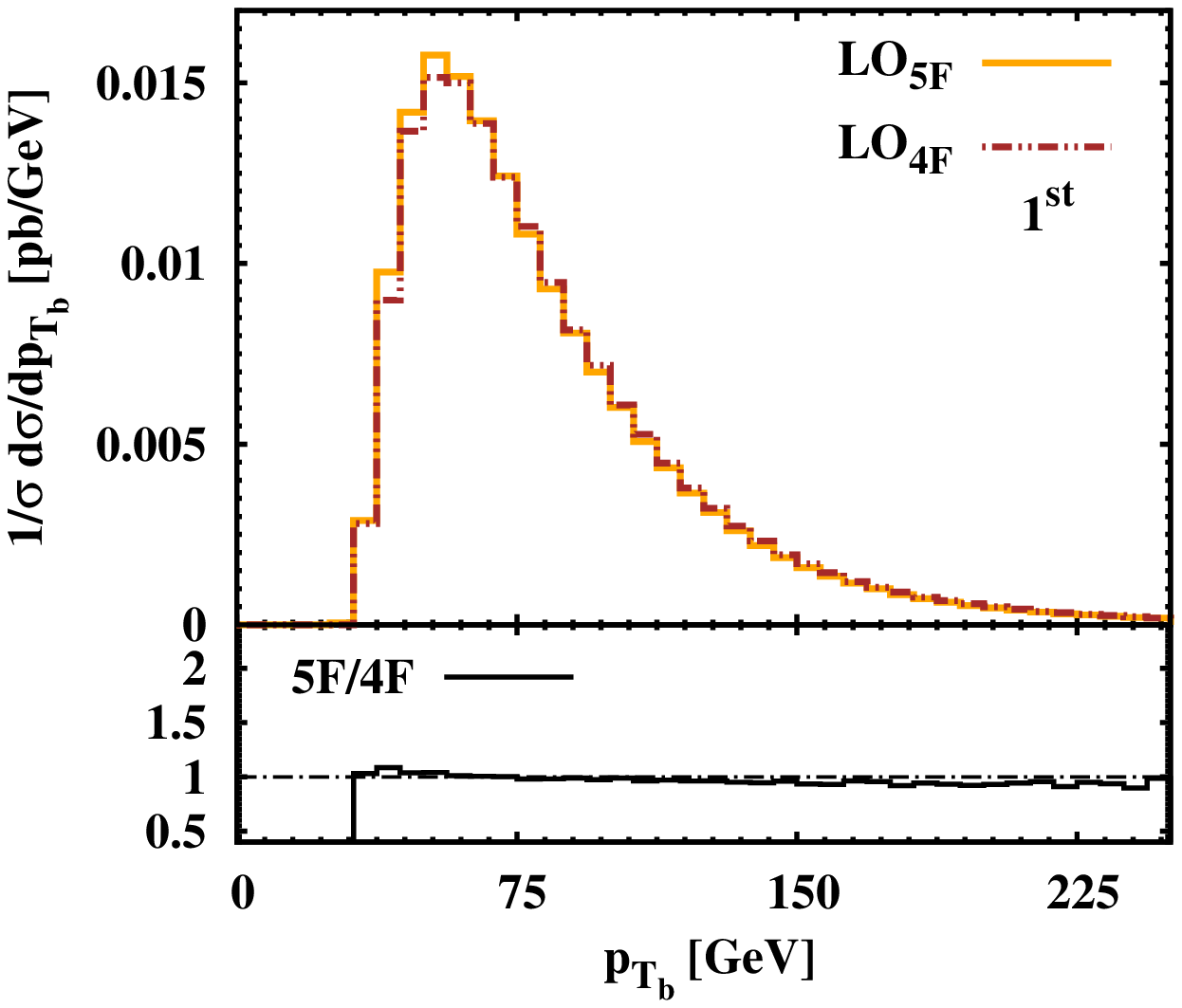}
\includegraphics[width=0.49\textwidth]{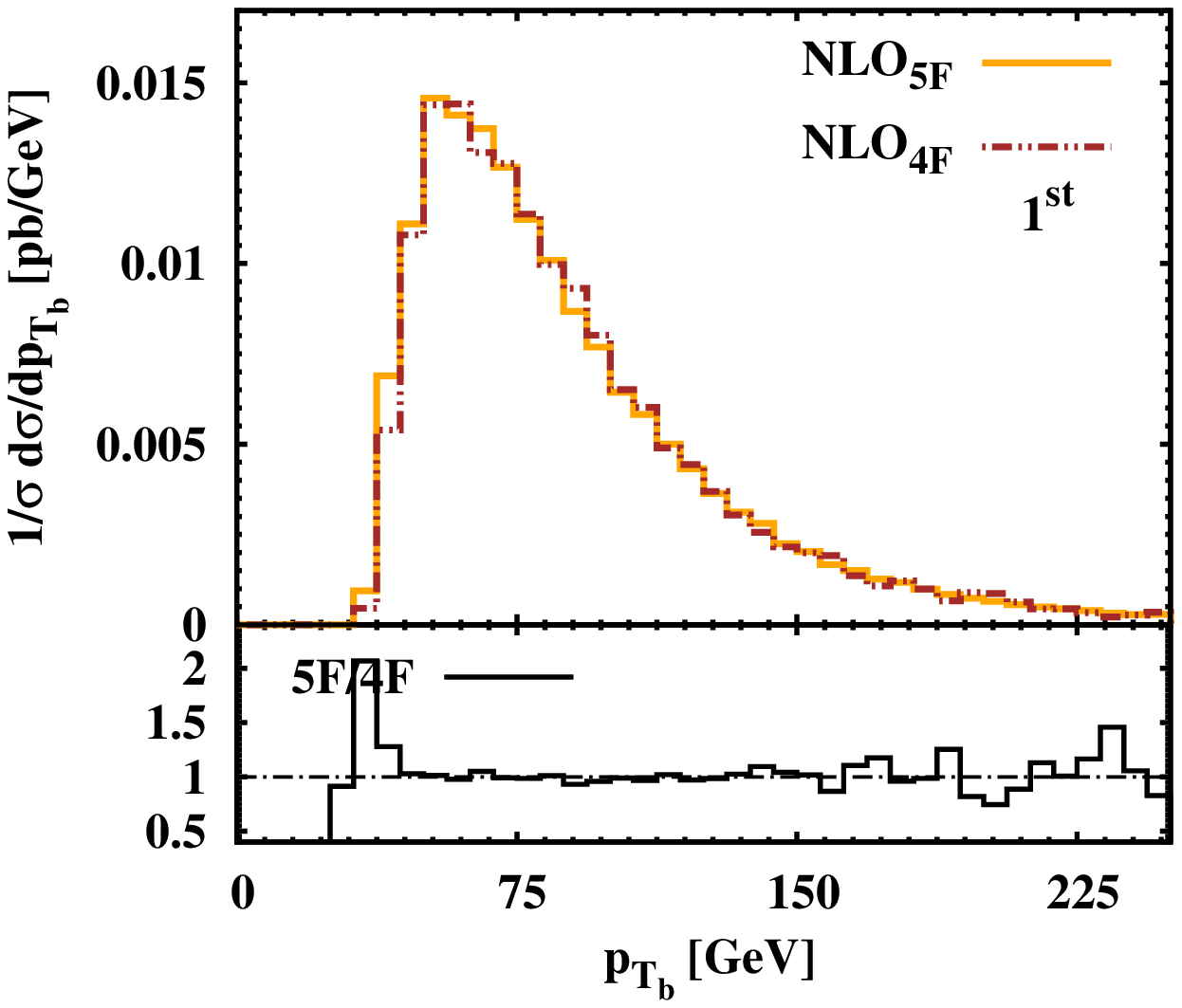}
\end{center}
\caption{\it \label{fig:nlo2:4f} Differential cross section for
  $pp\rightarrow b\bar{b} b\bar{b} ~+ X$ at the LHC ($\sqrt{s}$ = 14
  TeV) in the 4FS and 5FS  as a function of the transverse momentum of
  the hardest bottom jet. Also shown are the  normalised distributions
  at LO (lower left panel) and at  NLO (lower right panel).   The
  lower panels show the ratio of the results within the two schemes.
  The scale choice is $\mu_R = \mu_F = \mu_0 = H_T$, the cross
  sections  are evaluated with the 5FS and 4FS MSTW2008 pdf sets,
  respectively.}
\end{figure}

In Figure\,\ref{fig:nlo2:4f} we present the differential distribution
in the transverse momentum of the hardest bottom jet, as calculated in
the 5FS with massless bottom quarks and in the 4FS with  $m_b =
4.75$\,GeV. We show the absolute prediction at LO and NLO, and the
predictions normalised to the corresponding inclusive cross
section. The latter plots reveal that the difference in the shape of
the distributions in the 5FS and the 4FS is very small. We find
similar results for other differential distributions.

Let us finally present the comparison of the massive bottom quark
results as obtained with the Catani-Seymour and Nagy-Soper subtraction
schemes,  see Table\,\ref{tab:ffsns}. We observe full agreement
between the two calculations within the numerical error of the Monte
Carlo integration, and thereby validate our implementation of the NS
subtraction scheme also for the case of massive fermions. 
\begin{table}[h!]
\renewcommand{\arraystretch}{1.5}
\begin{center}
  \begin{tabular}{c|c|c|c}
\hline
$pp\to b\bar{b}b\bar{b}+X$  
& $\sigma_{\rm NLO}^{\rm CS \, (\alpha_{max}=1)}$ [pb] & 
$\sigma_{\rm NLO}^{\rm CS \, (\alpha_{max}=0.01)}$ [pb] & 
$\sigma_{\rm NLO}^{\rm NS}$ [pb] \\ \hline \hline 
MSTW2008NLO  (4FS) & $118.3 \pm 0.5$ 
& $118.2 \pm 0.7$ & $118.0 \pm$ 0.5 \\
\hline
  \end{tabular}
\end{center}
  \caption{\it \label{tab:ffsns} 4FS NLO cross sections for
$pp\rightarrow b\bar{b} b\bar{b} ~+ X$ at the LHC ($\sqrt{s}$ = 14
TeV). Results are shown for two different subtraction schemes, the
Catani-Seymour (CS) dipole subtraction, without ($\alpha_{\rm max}=1$)
and with ($\alpha_{\rm max}=0.01$) restriction on the phase space of
the subtraction, and the new Nagy-Soper (NS) scheme, including the
numerical error from the Monte Carlo integration. The renormalisation
and factorisation scales have been set to the central value $\mu_0 =
H_T$,  and the MSTW2008NLO 4FS pdf set has been employed.} 
 \end{table}


\subsection{Comparison with results presented in the literature}
%

A detailed comparison of our results with  Ref.\,\cite{Greiner:2011mp} has been
performed. We find agreement for the virtual amplitude at one specific phase space
point, but cannot reproduce the published results for the integrated
hadronic LO and NLO cross sections with the setup as described in
Ref.\,\cite{Greiner:2011mp}.  When the CTEQ6.5 pdf set \cite{Tung:2006tb} is
used rather than CTEQ6M \cite{Pumplin:2002vw} 
as specified in \cite{Greiner:2011mp}, and the factorization and renormalization 
scales are set to the common value 
\begin{equation}
\label{eq:scale}
\mu_0 = \frac{1}{4}\sqrt{\sum_{i} p^2_{T,\,i}}\,, 
\end{equation}
the LO
result published in \cite{Greiner:2011mp} can be reproduced as shown in Table \ref{tab:lo-comparison}:
\begin{table}[h!]
\renewcommand{\arraystretch}{1.5}
\begin{center}
  \begin{tabular}{c|c}
\hline
  $\sigma_{\rm LO}^{\mbox{\footnotesize{ \cite{Greiner:2011mp}}}}$ [pb] &  
$\sigma_{\rm LO}$ [pb] \\
\hline\hline
94.88 $\pm$ 0.14&
94.74 $\pm$ 0.20 
 \\
\hline
  \end{tabular}
\end{center}
  \caption{\it \label{tab:lo-comparison}  LO cross section for $pp
    \to b\bar{b}b\bar{b}$ + X at the LHC ($\sqrt{s}$ = 14 TeV) in
    comparison with the result of Ref.\,\cite{Greiner:2011mp}.
    Results are shown including the numerical error from the Monte
    Carlo integration. The scale choice is $\mu_R = \mu_F = \mu_0$,
    with $\mu_0$ as defined in Eq.\,(\ref{eq:scale}), and the cross sections are evaluated with the CTEQ6.5 pdf set.}
 \end{table}

After comprehensive numerical checks and the communication with the authors of Ref.\,\cite{Greiner:2011mp} it turned out that the 
NLO numbers published in \cite{Greiner:2011mp} are based on a scale setting that mixes partons and jets and that 
is not consistent with what is specified in Ref.\,\cite{Greiner:2011mp}. Adopting 
$\mu_R = \mu_F = \mu_0$, with $\mu_0$ defined in Eq.\,(\ref{eq:scale}), and summing over the transverse 
momenta of all jets, we obtain the NLO numbers presented in Table \ref{tab:nlo-comparison}. Note that 
at NLO the final state can consist of four or five jets, as determined by the jet algorithm. 
Our results are compared to the NLO number as published in Ref.\,\cite{Greiner:2011mp}, $\sigma_{\rm NLO}^{\mbox{ \footnotesize{\cite{Greiner:2011mp}}}}$, 
and to a corrected number obtained by means of private communication from the authors of Ref.\,\cite{Greiner:2011mp}, $\sigma_{\rm NLO}^{\mbox{\footnotesize{ \cite{Greiner:2011mp}, corr.}}}$. The corrected result agrees with our calculation. 

\begin{table}[h!]
\renewcommand{\arraystretch}{1.5}
\begin{center}
  \begin{tabular}{cc|ccc}
\hline
 $\sigma_{\rm NLO}^{\mbox{\footnotesize{ \cite{Greiner:2011mp}}}}$ [pb] &  $\sigma_{\rm NLO}^{\mbox{\footnotesize{ \cite{Greiner:2011mp}, corr.}}}$ [pb] &
$\sigma_{\rm NLO}^{\rm CS \, (\alpha_{max}=0.01)}$ [pb] 
& $\sigma_{\rm NLO}^{\rm CS \, (\alpha_{max}=1)}$ [pb] &
$\sigma_{\rm NLO}^{\rm NS}$ [pb] \\
\hline\hline
140.48 $\pm$ 0.64 & $143.75 \pm 0.67$ &
143.70$\pm$	0.44
&144.35 $\pm$ 0.53 &
144.73 $\pm$ 0.62
 \\
\hline
  \end{tabular}
\end{center}
  \caption{\it \label{tab:nlo-comparison}  NLO cross sections for $pp
    \to b\bar{b}b\bar{b}$ + X at the LHC ($\sqrt{s}$ = 14 TeV)  in
    comparison with published results of Ref.\,\cite{Greiner:2011mp}, $\sigma_{\rm NLO}^{\mbox{\footnotesize{ \cite{Greiner:2011mp}}}}$, and a corrected result based on the calculation of Ref.\,\cite{Greiner:2011mp}, $\sigma_{\rm NLO}^{\mbox{\footnotesize{ \cite{Greiner:2011mp}, corr.}}}$ (private communication).  Our results are shown for two different subtraction schemes, the
    Catani-Seymour (CS) dipole subtraction, without ($\alpha_{\rm
      max}=1$) and with ($\alpha_{\rm max}=0.01$) restriction on the
    phase space of the subtraction, and the new Nagy-Soper (NS)
    scheme, including the numerical error from the Monte Carlo
    integration. The scale choice is $\mu_R = \mu_F = \mu_0$, 
    with $\mu_0$ as defined in Eq.\,(\ref{eq:scale}), and 
        the
    cross sections are evaluated with the CTEQ6.5 pdf set.}
 \end{table}
Also note that in Ref.\,\cite{Greiner:2011mp} an NLO pdf set has been used
both for the LO and NLO result.  Although this may be justified to study 
the impact of higher-order corrections to the partonic cross section, it can be misleading when establishing the genuine effect
of higher-order corrections, the adequacy of the scale choice
or the size/shape of integrated/differential $K$-factors used in
experimental analyses when comparing Monte Carlo simulations to the
LHC data. 

\section{Summary}\label{sec:conclusion}
The
production of four bottoms quarks, $pp\rightarrow b\bar{b} b\bar{b}
~+X$, provides an important background for new physics searches at the
LHC. We have performed a calculation of the NLO QCD corrections to
this process with the \textsc{Helac-NLO} framework, employing a new
subtraction scheme for treating real radiation corrections at NLO
implemented in \textsc{Helac-Dipoles}.  Results have been presented for
inclusive and differential cross-sections for $pp \to b \bar{b} b
\bar{b} + X$ at the LHC at the centre-of-mass energy of ${\sqrt{s} =
14}$\,TeV. We find that the higher-order corrections significantly
reduce the scale dependence, with a residual theoretical uncertainty
of about 30\% at NLO. The impact of the bottom quark mass is moderate
for the cross section normalisation and negligible for the shape of
distributions. The fully differential NLO cross section calculation
for the process $pp\rightarrow b\bar{b} b\bar{b} ~+X$ presented in
this paper provides an important input for the experimental analyses
and the interpretation of new physics searches at the LHC.

\acknowledgments The calculations have been performed on the Grid
Cluster of the Bergische  Universit\"at Wuppertal, financed by the
Helmholtz-Alliance ``Physics at the  Terascale'' and the BMBF.  We
acknowledge the correspondence  with Nicolas Greiner on the results
of Ref.\,\cite{Greiner:2011mp} and discussions with  Zoltan Nagy. This
research was supported in part by the German Research Foundation (DFG)
via the Sonderforschungsbereich/Transregio SFB/TR-9 ``Computational
Particle Physics''.  The work of M.C. was supported by the DFG
Heisenberg program.  M.W. acknowledges support by the DFG under Grant
No. WO 1900/1-1 (``Signals and Backgrounds Beyond Leading
Order. Phenomenological studies for the LHC'').

 \end{document}